\documentclass[12pt,preprint]{aastex}

\newcommand{\myemail}{terada@subaru.naoj.org}

\slugcomment{Received 2012 April 21; accepted 2012 October 3}

\shorttitle{AO Observations for Water Ice in Silhouette Disks}
\shortauthors{Terada et al.}

\begin{document}

\title{ADAPTIVE OPTICS OBSERVATIONS OF 3\,$\micron$ WATER ICE IN SILHOUETTE DISKS IN THE ORION NEBULA CLUSTER AND M43}

\author{Hiroshi Terada\altaffilmark{1},
		Alan T. Tokunaga\altaffilmark{2},
		Tae-Soo Pyo\altaffilmark{1},
		Yosuke Minowa\altaffilmark{1},
		Yutaka Hayano\altaffilmark{1},
		Shin Oya\altaffilmark{1},
		Makoto Watanabe\altaffilmark{3},
		Masayuki Hattori\altaffilmark{1},
		Yoshihiko Saito\altaffilmark{4},
		Meguru Ito\altaffilmark{5},
		Hideki Takami\altaffilmark{1},  and 
		Masanori Iye\altaffilmark{6,7,8}
}
\email{\myemail}

\altaffiltext{1}{Subaru Telescope, National Astronomical Observatory
of Japan, 650 North A'ohoku Place, Hilo, HI 96720, USA}
\altaffiltext{2}{Institute for Astronomy, University of Hawaii, 2680
Woodlawn Drive, Honolulu, HI 96822, USA}
\altaffiltext{3}{Department of Cosmosciences, Hokkaido University,
  Kita 10, Nishi 8, Kita-ku, Sapporo, Hokkaido 060-0810, Japan}
\altaffiltext{4}{Department of Physics, Tokyo Institute of Technology, 
2-12-1 Ookayama, Meguro, Tokyo 152-8551, Japan}
\altaffiltext{5}{Department of Mechanical Engineering, University of Victoria, 
3800 Finnerty Road, Victoria, BC, V8P 5C2, Canada}
\altaffiltext{6}{National Astronomical Observatory of Japan,
  2-21-1 Osawa, Mitaka, Tokyo 181-8588, Japan}
\altaffiltext{7}{Department of Astronomy, University of Tokyo,
  7-3-1 Hongo, Bunkyo, Tokyo 113-0033, Japan}
\altaffiltext{8}{Department of Astronomical Science,
  The Graduate University for Advanced Studies (SOKENDAI), 
  2-21-1 Osawa, Mitaka, Tokyo 181-8588, Japan}

\begin{abstract}
We present the near-infrared images and spectra of four silhouette disks in the Orion Nebula Cluster (M42) and M43 using the Subaru Adaptive Optics system. While d053-717 and d141-1952 show no water ice feature at 3.1\,$\micron$, a moderately deep ($\tau_{ice}$ $\sim$ 0.7) water ice absorption is detected toward d132-1832 and d216-0939. 
Taking into account the water ice so far detected in the silhouette disks, the critical inclination angle to produce a water ice absorption feature is confirmed to be 65--75\arcdeg. As for d216-0939, the crystallized water ice profile is exactly the same as in the previous observations taken 3.63 years ago. If the water ice material is located at 30~AU, then the observations suggest it is uniform at a scale of about 3.5~AU.
\end{abstract}
\keywords{dust, extinction -- evolution -- infrared: ISM -- infrared: planetary systems -- protoplanetary disks -- stars: individual (d053-717, d132-1832, d141-1952, d216-0939)}

\section{Introduction}
A silhouette disk is one of the most suitable targets to investigate the primordial composition of a planetary disk by using its central star as a background light source. The Orion Nebula Cluster (ONC; M42) and its neighbor \ion{H}{2} region M43 are the nearest high-mass star-forming regions, exhibiting a number of silhouette disks. Since the first discovery by \citet{mcc96}, the number of the silhouette disks has significantly increased \citep{bal00,smi05,ric08} and currently the number of the pure silhouette disks is $\sim$30, which enables us to conduct a systematic study for disk absorption features. 

Water ice in protoplanetary disks is particularly important, because it is thought to evolve to icy bodies such as comets in our solar system. In theory, the distribution of the water ice in the protoplanetary disks has been extensively investigated as a key parameter for the planet formation \citep[e.g.,][]{hay81}, which predicts the inner boundary around 2~AU of the water ice presence (``snow line'')  where the temperature is 170K for the ice condensation. 
Although the mid-plane of the disks is of the greatest interest for the planet formation, it is difficult to observationally probe into the mid-plane of the disks due to high extinction. 
\citet{pon05} constructed the model of the water ice absorption in the circumstellar disk through the water ice detection for the edge-on disk object CRBR 2422-3423 in $\rho$ Ophiuchus dark cloud, which suggests a critical inclination angle of the disk for strong water ice detection to be around 69$\arcdeg$. Since silhouette disks show a clear morphology of the circumstellar disks, the inclination angle of the disks can be more accurately estimated. In fact, the silhouette disks located in M42 and M43 show a wide range of the disk inclination angle, and those are very suitable for examining the critical inclination angle in a larger sample. The silhouette disks in M42 and M43 are exposed to the intense UV radiation from O and B stars, and the critical inclination angle may be increased to a larger value by dissociation and desorption process of the outer layer of the disks. Since star formation occurs more commonly in high-mass star-forming regions, it is important to understand the water ice formation under such environments.

Strong water ice absorption at 3\,$\micron$ band is very useful to investigate characteristics of the water ice materials. In particular, it is well known that the profile and peak of the water ice absorption in the wavelengths of 2.9--3.3\,$\micron$ provides the information of its grain size and crystallinity. While larger water ice grains bring its absorption peak toward longer wavelength, more crystallinity causes a sharper peak at longer wavelength \citep{smi89}. The water ice feature so far detected at 3\,$\micron$ band toward young stellar objects is usually explained by the combination of amorphous and crystallized water ice absorption with a grain size of 0.1--0.5\,$\micron$, whose peaks are located at 3.0--3.05\,$\micron$ and $\sim$ 3.1\,$\micron$, respectively \citep{dar01}. 

\citet[][hereafter Paper I]{ter12} carried out $L$-band spectroscopy of five bright young stellar objects with silhouette disks in M42 and M43 and detected a deep water ice absorption ($\tau_{ice}$=0.67) toward d216-0939 in M43, which is associated with a large silhouette disk ($\sim$1000~AU in diameter). In the profile of the water ice absorption, the 3.2\,$\micron$ enhancement that is characteristics of crystallized water ice with large grains \citep[$\sim$1\,$\micron$;][]{leg79} was seen, and it suggested grain growth and thermal processing in the disk. Since the nearby star JW 671 \citep{jon88} shows no water ice feature, the water ice detected for d216-0939 is concluded to reside in the protoplanetary disk of d216-0939. Whereas d121-1925 also shows the water ice absorption, its nearby star JW 370 \citep{jon88} exhibits the same depth and profile of the water ice absorption, which suggests that the water ice toward d121-1925 is in the foreground. There is no evidence of the water ice feature detected in the remaining three objects: d182-332, d183-405, and d218-354. The low inclination angle of the disk is likely the primary cause for no water ice feature in the disk in the line of sight for d121-1925, d182-332, d183-405, and d218-354. Another possible cause is low abundance of the water ice in the disk due to the strong UV radiation from O and B stars. Increasing the number of spectra will help to understand the prevalence of water ice in the silhouette disks.

Adaptive optics (AO) with large ground-based telescopes gives us a unique opportunity to obtain the highest spatial resolution with diffraction limited images, which reaches to $<$ 0\farcs1 resolution at 3\,$\micron$. At the distance to M42 \citep[414 pc;][]{men07} and M43 \citep[440 pc;][]{ode08}, the direct study of the outer region of protoplanetary disks can be done with 40 AU resolution. Many circumstellar disks have been so far detected using AO, including the disks with high inclination angle ($>$80$\arcdeg$) which sometimes show an edge-on morphology \citep{mon00,jay02,per06}. In addition, AO observation has another significant advantage to provide increased sensitivity for photometry of point sources and use of a narrower slit for spectroscopy. It is crucial to use AO to increase the reachable targets associated with the silhouette disks. For AO observations, a nearby and bright reference source to sense the wavefront of light from the target region is required. For targets that do not have such nearby and bright natural stars, an artificial laser guide star (LGS) is available at some large telescopes. In the outskirt regions of M42 and M43, LGS plays an essential role to make the observation of silhouette disk possible.

Using the Subaru Telescope, we present natural guide star (NGS) and LGS images and spectra of four sources with silhouette disks in the outskirt of M42 and M43 regions, namely d053-717, d132-1832, d141-1952, and d216-0939 \citep{bal00,smi05}. In Section 2, our sample selection is explained. The observational procedures and applied data analysis are described in Section 3. In Section 4, the AO images and spectra for all the targets are shown, and the water ice optical depths are derived after the spectral continuum fit. The water ice feature in each disk is discussed in Section 5. The summary is described in Section 6.

\section{Samples}
\begin{figure}
\begin{center}
\includegraphics[scale=1.5]{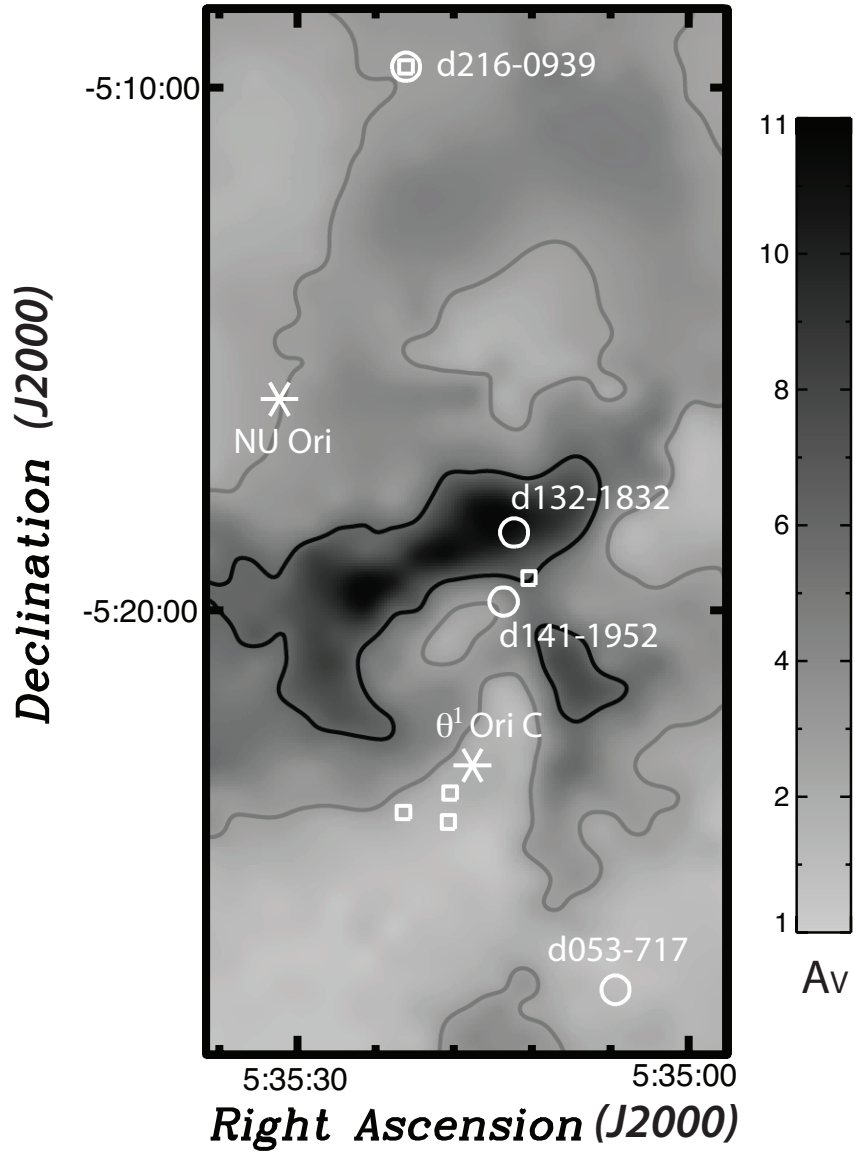}
\caption{\label{fig-sample}Location of our samples. Asterisks represent the exciting O and B stars, namely NU Ori for M43 and $\theta$$^{1}$ Ori C for M42. Open squares are the previous samples in Paper I, and open circles are corresponding to the samples in this paper. d216-0939 is observed in both runs. The samples are superposed on the visual extinction ($A_{V}$) map for the ONC members from Figure 6 of \citet{sca11}. Gray and black contours are at the $A_{V}$=3 and 6~mag, respectively. d053-717 is found to be located in the region of the least extinction ($A_{V}$ $<$ 3) among the four targets. d141-1952 and d216-0939 have moderate extinction ($A_{V}$ $\sim$ 3), and there is the largest extinction around the region of d132-1832 ($A_{V}$ $\sim$ 10).
}
\end{center}
\end{figure}

Targets were selected from all the pure silhouette disk samples \citep{smi05,ric08} with three criteria of (1) an inclination angle of $>$55$\arcdeg$, (2) a distance of $>$ 200$\arcsec$ from the O and B stars,  and (3) a brightness of $K^{\prime}$ $<$ 15.5. As for the first criterion about the inclination angle, Paper I found that the silhouette disks likely have the critical inclination angle of 65$\arcdeg$--75$\arcdeg$ for the detection of the water ice in disks, and it is required to confirm this possible critical angle by observing more targets with inclination angle 65$\arcdeg$--75$\arcdeg$. The second criterion is to avoid the environmental effect of strong UV radiation from the O and B stars. In fact, the previous samples with no water ice feature in Paper I are all close to the Trapezium cluster, and it was difficult to distinguish the environmental cause and the geometrical cause for no water ice detection. The third criterion is for obtaining sufficient signal-to-noise ratio (S/N) for spectroscopy at $L$. Typically, $L$ $\le$ 14 is required for detection of absorption feature with good S/N using an 8 m telescope. We selected the targets which met this criterion through a $K^{\prime}$ imaging survey (H. Terada et al., in preparation). In Figure~\ref{fig-sample}, the location of our targets is superposed on the extinction map derived by \citet{sca11}. Table~\ref{tbl-sample} summarizes the disk parameters of each target.

\begin{deluxetable}{c c c c c c}
	\tablecolumns{6}
	\tablewidth{0pt}
    \tablecaption{Disk Parameters\label{tbl-sample}}
	\tablehead{
      \colhead{}&\colhead{}&\colhead{}&\colhead{Inclination}&\colhead{Exciting}&\colhead{}\\
      \colhead{Object}&\colhead{Mass\tablenotemark{a}}&\colhead{Diameter}&\colhead{Angle}&\colhead{Source}&\colhead{Distance}\\
       \colhead{}&\colhead{($\times$10$^{-2}$ $M_{\sun}$)}&\colhead{(\arcsec)}&\colhead{(\arcdeg)}&&\colhead{(\arcsec)}
	}
	\startdata
       d053-717&0.52$\pm$0.07&0.9\tablenotemark{b}&65--85\tablenotemark{b}&$\theta^{1}$ Ori C&286.32\\
       d132-1832&0.80$\pm$0.08&1.5\tablenotemark{c}&75\tablenotemark{c}&$\theta^{1}$ Ori C&294.19\\
       d141-1952&1.50$\pm$0.06&0.7\tablenotemark{b}&55--60\tablenotemark{b}&$\theta^{1}$ Ori C&214.18\\
       d216-0939&4.53$\pm$0.06&2.6\tablenotemark{b}&75--80\tablenotemark{b}&NU Ori&400.49\\
	\enddata
	\tablenotetext{a}{\citet{man10}.}
	\tablenotetext{b}{\citet{smi05}.}
	\tablenotetext{c}{\citet{bal00}.}
	\tablecomments{Distance is measured from the listed O,B stars responsible for excitation of the \ion{H}{2} region.}
\end{deluxetable}

\section{Observations and Data Reduction}
All the observations were performed using Infrared Camera and Spectrograph \citep[IRCS:][]{tok98,kob00,ter04} combined with the Subaru Adaptive Optics system \citep[AO188;][]{hay10}. IRCS provides functions of imaging and spectroscopy with grisms and a cross-dispersed echelle in the wavelength range of 0.9--5.6\,$\mu$m, and it is optimized for the high angular resolution images delivered by the AO. The observing log is summarized in Table~\ref{tbl-obslog}, which includes information about the airmass of each observation and standard stars. 
The NGS mode was used for observations of d216-0939, since a sufficiently bright star ($R$$\sim$10~mag) for the wavefront sensing is located at 23\farcs42 north away from the object. For the other targets: d053-717, d132-1832, and d141-1952, the observations utilized the AO188 LGS \citep{hay10} for higher order wavefront correction together with a tip-tilt and defocus correction using an $R$$\sim$14~mag star. As for d053-717 and d141-1952, the targets themselves are bright enough ($R$=13.7~mag and $R$=13.6~mag, respectively) to be used as a reference star for the lower order of wavefront correction (tip-tilt guide star, TTGS). In the case of d132-1832 observations, the bright star ($R$=14.8~mag) located at 49$\arcsec$ southwest away from the object was used as TTGS.

Imaging was conducted by using a nine-point box-shaped dithering with an offset separation of 5\farcs0. Spectroscopy was performed using A-BB-A dithering sequence along the slit with an offset separation of 1\farcs5. The slit width was 0\farcs225, which corresponds to a spectral resolution ($\lambda$/$\Delta\lambda$) of 850 at $K$ and 510 at $L$. Sky flat was used for $L^{\prime}$, and the flat fielding for the $J$, $H$, and $K^{\prime}$ images and $K$ and $L$ spectra is done with an integration sphere illuminated by a halogen lamp. 

Standard analysis procedure is applied using the IRAF packages for all the data reduction of sky subtraction, flat fielding, shift-and-add and combining of the images and spectra, telluric correction, and wavelength calibration. FS 13 ($K^{\prime}$=10.126) and FS 154 \citep[$K^{\prime}$=11.050;][]{cas92} were used as $K^{\prime}$ photometric standard stars. For $L^{\prime}$ photometric calibration, observations of HD 40335 ($L^{\prime}$=6.441) and HD 1160 \citep[$L^{\prime}$=7.055;][]{leg03} were conducted. For correction of the telluric absorption, HR 2075 (A0IV), HR 2328 (A0Vn), and HR 1826 (A0Vn) were observed with an assumption that those stars have a blackbody continuum with a temperature of 9480 K. The hydrogen absorption of these spectroscopic standard stars is appropriately removed using the method developed by \citet{vac03}. The Br $\gamma$ emission at 2.166\,$\micron$ is likely the result of poor subtraction of the nebular emission and is not intrinsic to the source. The airmass mismatch between the object and the standard observations was very small as shown in Table~\ref{tbl-obslog}. Wavelength calibration was done using the telluric absorption lines in the spectrum. 

\begin{deluxetable}{c c c c c c c c c}
	\tabletypesize{\scriptsize}
    \tablecolumns{9}
    \tablewidth{0pt}
    \tablecaption{Observing log\label{tbl-obslog}}
    \tablehead{
      \colhead{}&\colhead{Observing}&\colhead{}&\colhead{Exposure}&\colhead{Average}&\colhead{Standard}&\colhead{$\Delta$Airmass}&\colhead{PSF}&\colhead{}\\
      \colhead{Object}&\colhead{Date}&\colhead{Mode}&\colhead{Time}&\colhead{Airmass}&\colhead{Star}&\colhead{Obj.-Std.}&\colhead{Reference Star}&\colhead{Offset}\\
      \colhead{}&\colhead{(UT)}&\colhead{}&\colhead{(s)}&\colhead{}&\colhead{}&\colhead{}&\colhead{}&\colhead{(\arcsec)}}
      \startdata
      d053-717&2011 Oct 23&$K^{\prime}$ imaging&15&1.115&FS 13&$-$0.040&JW 283&10.79\\
      &2011 Oct 23&$L^{\prime}$ imaging&50&1.118&HD 40335&$-$0.036&&\\
      &2011 Oct 23&$K$ spectroscopy&120&1.113&HR 2075&$-$0.026&&\\
      &2011 Oct 23&$L$ spectroscopy&480&1.108&HR 2075&$-$0.015&&\\
      d132-1832&2011 Oct 23&$K^{\prime}$ imaging&90&1.164&FS 13&+0.008&2MASS J05351361-0518329&5.68\\
      &2011 Oct 23&$L^{\prime}$ imaging&100&1.173&HD 40335&+0.020&&\\
      &2011 Oct 24&$K$ spectroscopy&480&1.175&HR2328&+0.047&&\\
      &2011 Oct 24&$L$ spectroscopy&3240&1.119&HR 2328,&+0.031&&\\
      d141-1952&2011 Oct 23&$K^{\prime}$ imaging&3&1.138&FS 13&$-$0.018&JW 413&7.04\\
      &2011 Oct 23&$L^{\prime}$ imaging&50&1.140&HD 40335&$-$0.013&&\\
      &2011 Oct 23&$K$ spectroscopy&60&1.139&HR 2075&$-$0.001&&\\
      &2011 Oct 23&$L$ spectroscopy&120&1.127&HR 2075&$-$0.0001&&\\
      d216-0939&2008 Sep 9&$J$ imaging&240&1.175&FS 154&$-$0.005&JW 671&4.38\\
      &2008 Sep 9&$H$ imaging&270&1.161&FS 154&$-$0.024&&\\
      &2008 Sep 9&K imaging&270&1.194&FS 154&+0.007&&\\
      &2008 Oct 9&$L^{\prime}$ imaging&560&1.322&HD 1160&+0.070&&\\
      &2009 Oct 2&$L$ spectroscopy&1000&1.104&HR 1826&$-$0.007&&\\
      \enddata
\end{deluxetable}

\section{Results}
\subsection{1.2--4.2\,$\micron$ Images with Adaptive Optics}
Aperture photometry is applied for all the images with a radius of 1\farcs5, and the photometric results are summarized in Table~\ref{tbl-phot}. These photometric values include the scattered light component around the central star.

Images of the targets are presented in Figure~\ref{fig-objimage}, which show good image quality with FWHM of $\le$0\farcs2. Table~\ref{tbl-str} shows the achieved spatial resolution and Strehl ratio for the images. The Strehl ratio is calculated from the peak intensity normalized by that of point-spread-function (PSF) without any turbulence. To investigate possible extended features around the sources, subtracted images by their nearby bright stars are presented in the bottom panels of Figure~\ref{fig-objimage}. The reference stars for the PSF are chosen at a vicinity of the objects. JW 283, 2MASS J05351361-0518329, JW 413, and JW 671 are located at 10\farcs79 northeast, 5\farcs68 east, 7\farcs04 west, and 4\farcs38 southwest away from each object, respectively. Whereas it is well known that the PFS obtained with an AO system is affected by the anisoplanatism, the Strehl ratio degradation for the separation of 10$\arcsec$ is only up to $\sim$ 0.1 by measuring with the AO188 system in both NGS mode \citep{min10} and LGS mode \citep{min12}. The anisoplanatism for the PSF reference stars with an offset of $\leq$ 10\farcs79 does not affect the following analysis. In the subtracted images, extended features are seen around $K^{\prime}$ image of d132-1832 and $J$, $H$, $K^{\prime}$, and $L^{\prime}$ images of d216-0939. As for the images of d141-1952, the reference star is found to be extended, which is known to be a young stellar object with a fluorescent Fe line \citep{tsu05}. As a result, a negative extension appears at both $K^{\prime}$ and $L^{\prime}$. 

The extended feature at $K^{\prime}$ of d132-1832 is elongated along the position angle of the silhouette disk. In geometry of nearly edge-on disks, the dark lane is usually seen between two sides of scattered light from the central star. In fact, the $Hubble$ $Space$ $Telescope$ ($HST$) Advanced Camera for Surveys (ACS) image at $Ic$, shown in \citet{ric08}, marginally exhibits such morphology. It is probable that the spatial resolution of 0\farcs2 is not enough to resolve the possible dark lane of this object in the infrared. Also, no outflow signature of molecular hydrogen emission which is seen in our spectrum of d132-1832 is exhibited in the subtracted image at $K^{\prime}$. The object d216-0939 is known to have a large reflection nebula extended by $\sim$3$\arcsec$ and was discovered in the optical using $HST$ ACS by \citet{smi05}. The infrared extended feature detected at $J$, $H$, and $K^{\prime}$ resembles the morphology in the optical. Regarding the reflection nebula at $L^{\prime}$, it is detected on only the east side of the dark lane and the surface brightness is much fainter than that detected in the $J$, $H$, and $K^{\prime}$ images.

\begin{figure*}
\begin{center}
\includegraphics[scale=1]{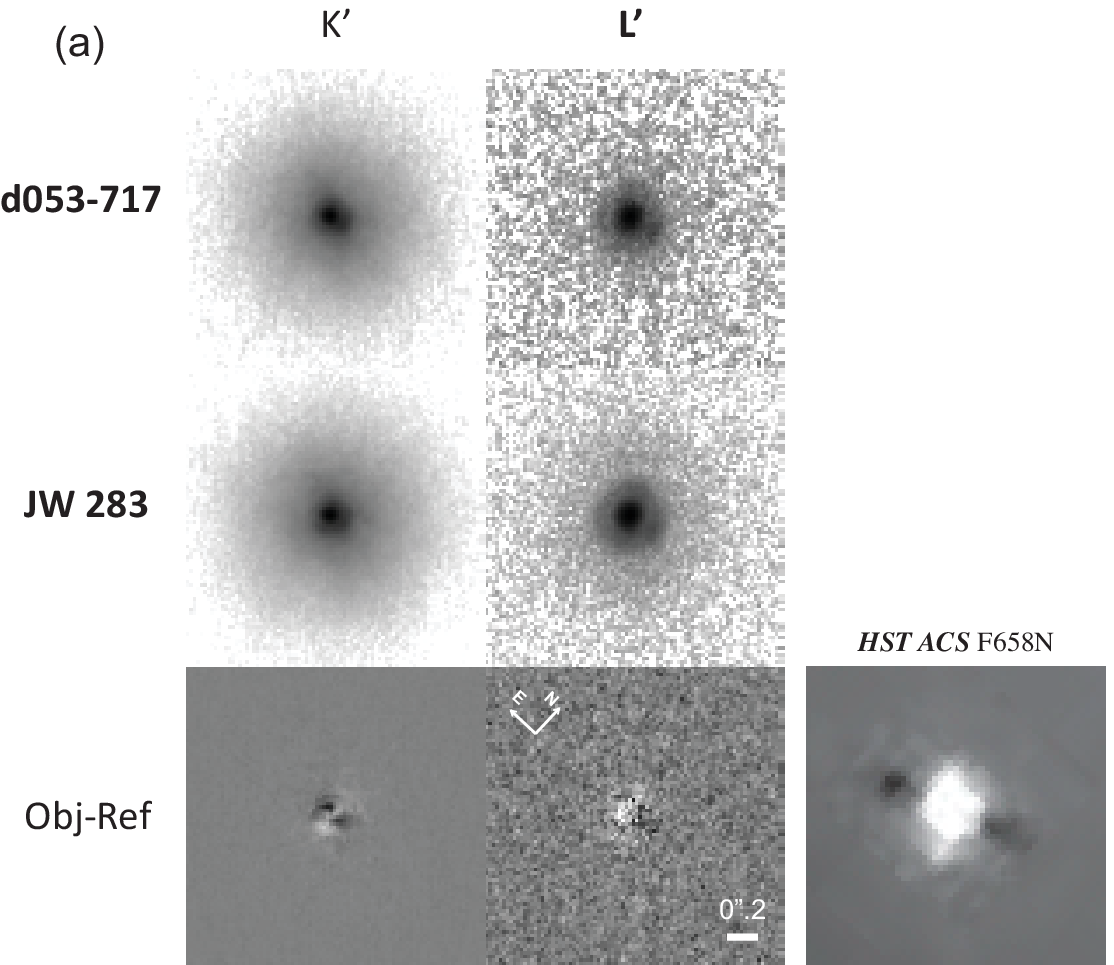}
\caption{\label{fig-objimage}Adaptive optics images of the four targets with the silhouette disks. For each figure (a)--(d), the top panel shows the object image, and the PSF reference star is presented in the middle panel. Images in the top and middle panels are normalized by a peak flux and those are displayed in logarithmic scale with black for 100\% and white for 0.2\% of the peak flux. The bottom panel for $J$, $H$, $K^{\prime}$, and $L^{\prime}$ shows the subtracted image of the objects by the PSF reference. This figure is displayed in linear scale with black for 10\% and white for $-$10\% of the peak flux. 
For comparison, the bottom-right panel shows the $HST$ ACS H$\alpha$ image from \citet{smi05}. All the images have the same size of 1\farcs75 $\times$ 1\farcs75.
}
\end{center}
\end{figure*}

\begin{figure*}
\begin{center}
\figurenum{2}
\includegraphics[scale=1]{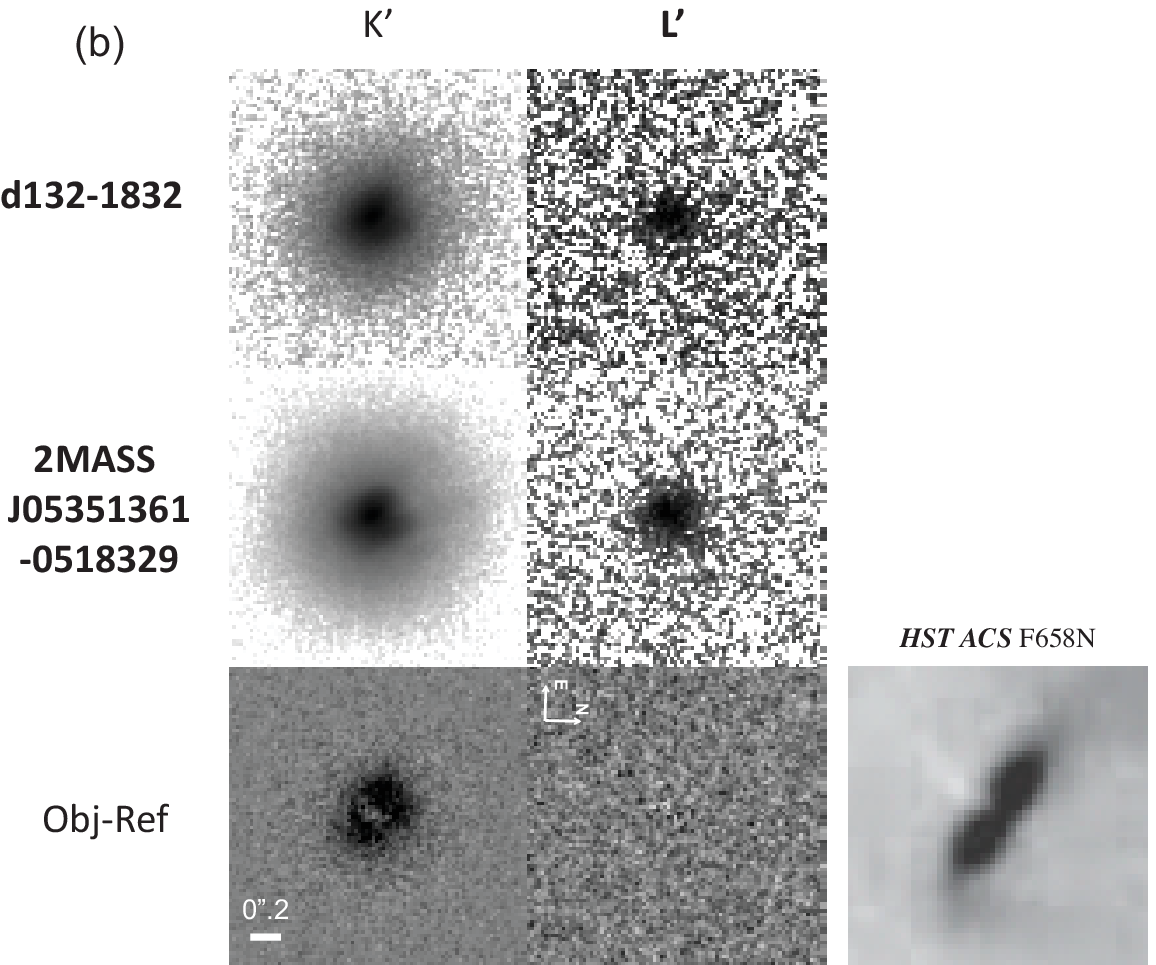}
\caption{\it Continued.
}
\end{center}
\end{figure*}

\begin{figure*}
\begin{center}
\figurenum{2}
\includegraphics[scale=1]{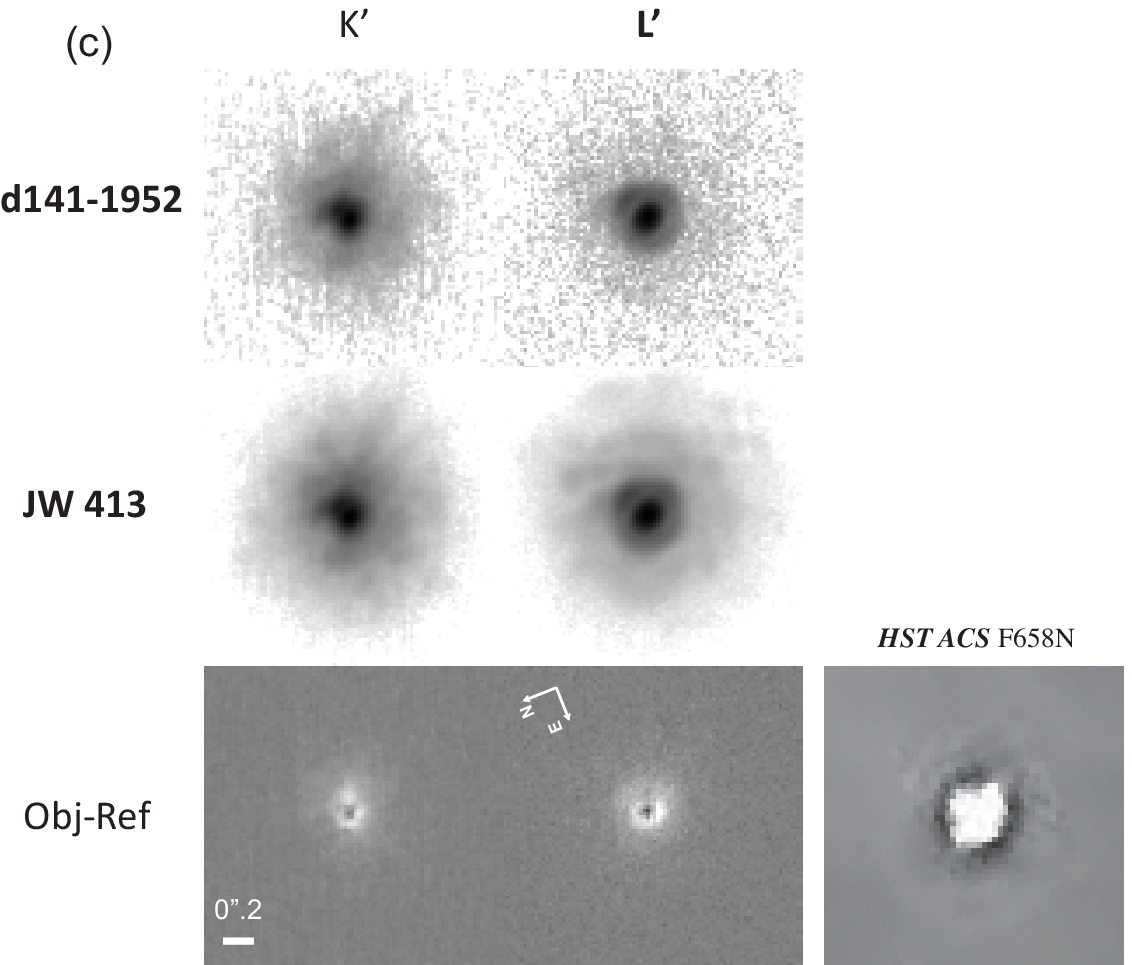}
\caption{\it Continued.
}
\end{center}
\end{figure*}

\begin{figure*}
\begin{center}
\figurenum{2}
\includegraphics[scale=1]{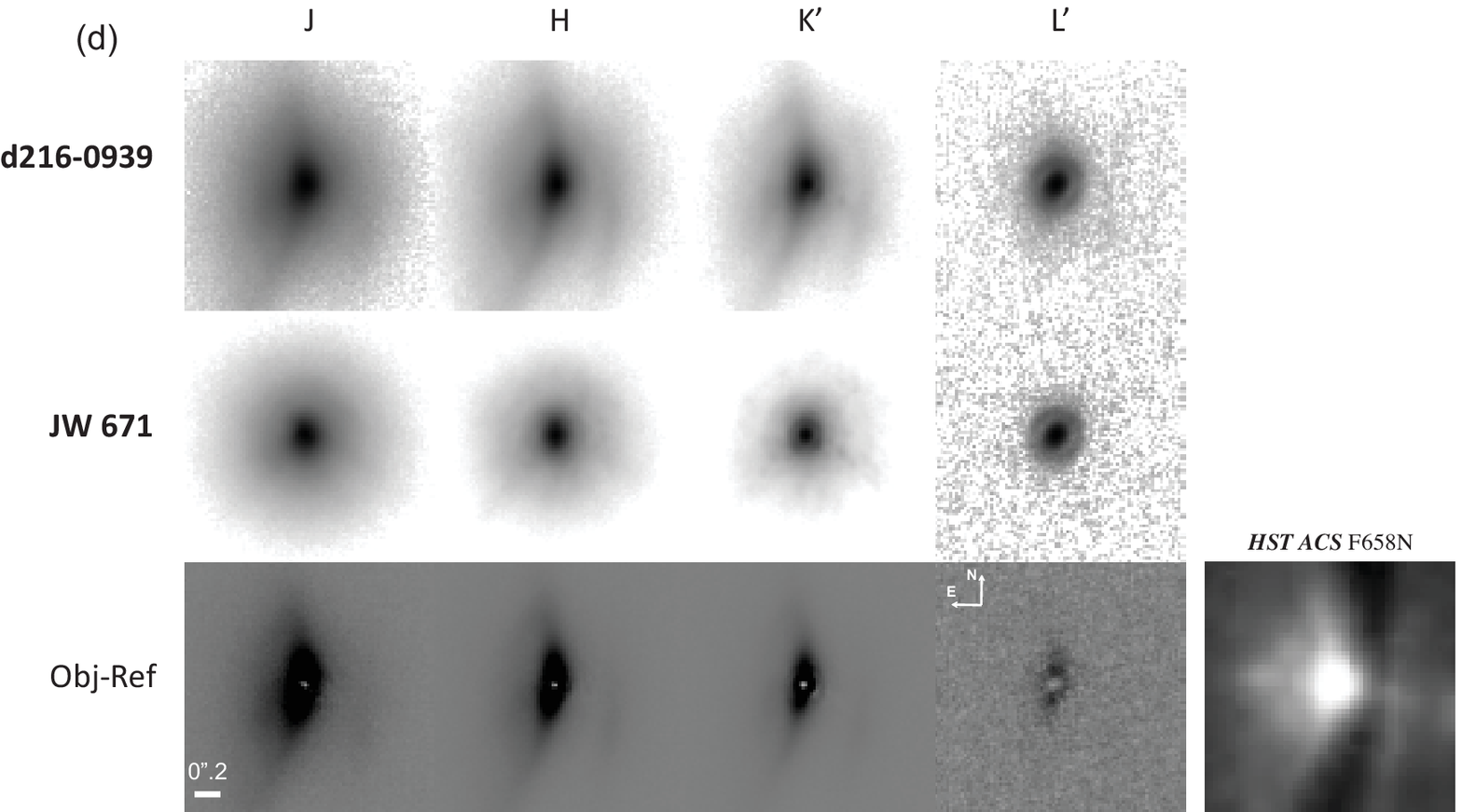}
\caption{\it Continued.
}
\end{center}
\end{figure*}

\begin{deluxetable}{c c c c c}
    \tablecolumns{5}
    \tablewidth{0pt}
    \tablecaption{Total Magnitudes\label{tbl-phot}}
    \tablehead{
      \colhead{Object}&\colhead{$J$}&\colhead{$H$}&\colhead{$K^{\prime}$}&\colhead{$L^{\prime}$}}
      \startdata
      d053-717&\nodata&\nodata&10.81&11.12\\
      d132-1832&\nodata&\nodata&15.06&13.62\\
      d141-1952&\nodata&\nodata&10.20&9.53\\
      d216-0939&14.43&13.58&12.35&10.74\\
      \enddata
      \tablecomments{The typical 1 $\sigma$ uncertainties are 0.02, 0.02, 0.02, and
      0.1 mag for $J$, $H$, $K^{\prime}$, and $L^{\prime}$, respectively.}
\end{deluxetable}

\begin{deluxetable}{c c c c c}
    \tablecolumns{4}
    \tablewidth{0pt}
    \tablecaption{Achieved Spatial Resolution and Strehl Ratio\label{tbl-str}}
    \tablehead{
      \colhead{Object}&\colhead{Band}&\colhead{FWHM}&\colhead{Strehl}\\
      \colhead{}&\colhead{}&\colhead{(\arcsec)}&\colhead{Ratio}}
      \startdata
      d053-717&$K^{\prime}$&0.092&0.12\\
      &$L^{\prime}$&0.11&0.44\\
      d132-1832&$K^{\prime}$&0.20&0.05\\
      &$L^{\prime}$&0.16&0.30\\
      d141-1952&$K^{\prime}$&0.11&0.13\\
      &$L^{\prime}$&0.12&0.48\\
      d216-0939&$J$&0.13&0.035\\
      &$H$&0.11&0.086\\
      &$K^{\prime}$&0.081&0.25\\
      &$L^{\prime}$&0.12&0.55\\
      \enddata
\end{deluxetable}

\subsection{1.9--4.2\,$\micron$ Spectra with Adaptive Optics}
\begin{figure}
\begin{center}
\includegraphics[scale=0.7]{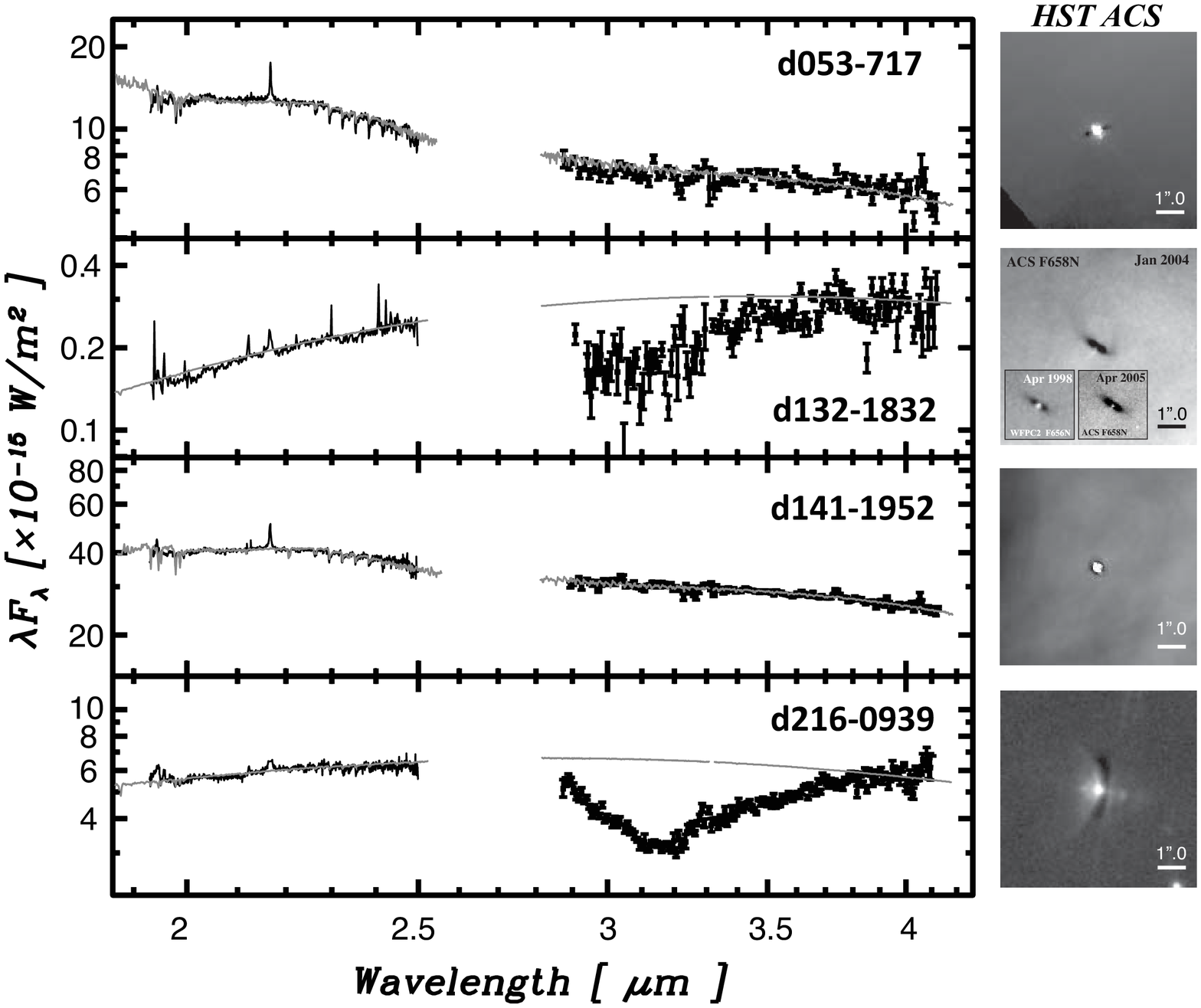}
\caption{\label{fig-objspec}Spectra of the four targets associated with the silhouette disks. Black lines show the spectra of the targets, and gray lines are the model spectra. The absolute level of the spectrum is determined from the flux, which falls in the slit. In the right panel, $HST$ ACS H$\alpha$ images are shown with a size of 7$\arcsec$ $\times$ 7$\arcsec$ from \citet{smi05}. 
Only the d132-1832 image in 2005 April shown as one of the inserts is the online data presented in \citet{ric08}.
}
\end{center}
\end{figure}

Figure~\ref{fig-objspec} shows the spectra of the four targets: d053-717, d132-1832, d141-1952, and d216-0939. The level of the continuum at $K$ and $L$ is determined by estimating the flux in the slit taken from the images shown in Figure~\ref{fig-objimage}. The same procedure for the spectral model fitting as described in Paper I is applied for all the spectra using IRTF spectral library \citep{cus05,ray09}, and the best-fit model spectra are overplotted with gray lines in the same panel of Figure~\ref{fig-objspec}. The best-fit parameters are shown together with the previously known spectral type and $A_{V}$ in Table~\ref{tbl-fitpara}. It is noted that the derived $A_{V}$ could be underestimated due to the scattered light around the central sources.

\begin{deluxetable}{c c c c c c c}
	\tablecolumns{7}
	\tablewidth{0pc}
	\tablecaption{Spectral Type, Visual Extinction, and Blackbody Temperature\label{tbl-fitpara}}
      \tablehead{
      \colhead{}         &\multicolumn{2}{c}{\citet{hil97}}&\colhead{}&\multicolumn{3}{c}{Derived Parameter}\\
      \cline{2-3} \cline{5-7}\\
      \colhead{}&\colhead{Spectral}&\colhead{}&\colhead{}&\colhead{Spectral}&\colhead{}&\colhead{Blackbody}\\
      \colhead{Object}&\colhead{Type}&\colhead{$A_{V}$}&\colhead{}&\colhead{Type}&\colhead{$A_{V}$}&\colhead{Temperature}\\
      \colhead{}&\colhead{}&\colhead{(mag)}&\colhead{}&\colhead{}&\colhead{(mag)}&\colhead{(K)}}
      \startdata
       d053-717&K5--6&0.43&&M2$\pm$1&1.5$\pm$0.20&1100\\
       d132-1832&\nodata&\nodata&&K5$\pm$2&1.5$\pm$0.30&1050\\
       d141-1952&Late G&2.55&&M2.5$\pm$1&2.55$\pm$0.15&1100\\
       d216-0939&K5&0.72&&K5$\pm$1&0.72$\pm$0.18&1190\\
      \enddata
\end{deluxetable}

\begin{figure}
\begin{center}
\includegraphics[scale=0.7]{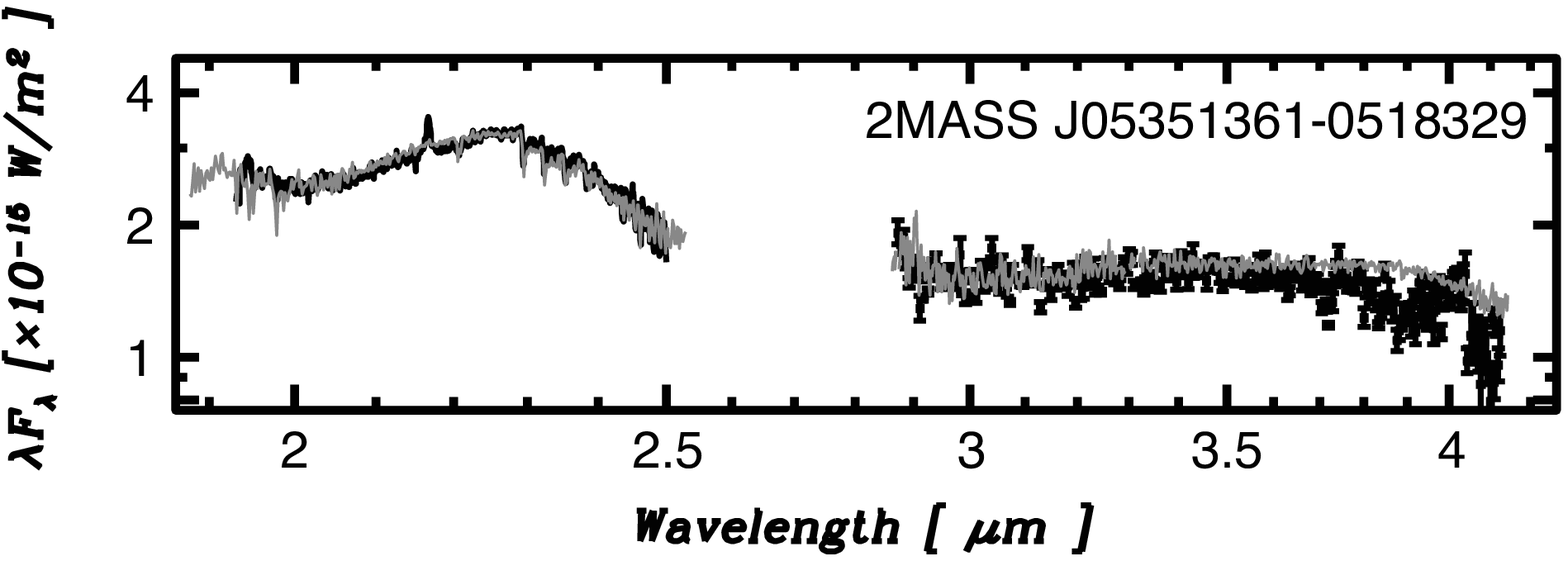}
\caption{\label{fig-refspec}Spectrum of the star 2MASS J05351361-0518329 which is near d132-1832. The model fit is shown as a gray line. No apparent water ice feature is detected in the spectrum.
}
\end{center}
\end{figure}

A broad absorption feature around 3.1\,$\micron$ is seen in the spectrum of d132-1832 and d216-0939. On the other hand, there are no absorption signature at 3.1\,$\micron$ band for d053-717 and d141-1952. The center wavelengths, optical depths, and band-widths of this absorption feature for each target are derived from the spectra (Table~\ref{tbl-tau}). Whereas S/N for d132-1832 is relatively low ($\sim$10 at the continuum) due to its faintness at $L^{\prime}$ (13.62~mag), the profile of the 3.1\,$\micron$ absorption matches the water ice band. In contrast, the nearby star 2MASS J05351361-0518329 shows no water ice feature in the spectrum (Figure~\ref{fig-refspec}) based on the best-fit model spectra with a spectral type of M7V, $A_{V}$=6~mag, and a blackbody temperature of 1000 K. 

\begin{deluxetable}{c c c c}
	\tablecolumns{4}
	\tablewidth{0pc}
    \tablecaption{Optical Depth of Water Ice Absorption at 3.1\,$\micron$\label{tbl-tau}}
	\tablehead{
      \colhead{Object}&\colhead{Peak}&\colhead{$\tau_{ice}$}&\colhead{$\Delta\nu$}\\        	\colhead{}&\colhead{($\micron$)}&\colhead{}&\colhead{(cm$^{-1}$)}
	}
      \startdata
      d053-717&\nodata&$\le$0.065&\nodata\\
      d132-1832&3.08&0.70$\pm$0.05&445\\
      d141-1952&\nodata&$\le$0.0052&\nodata\\
      d216-0939&3.12&0.74$\pm$0.02&472\\
      \enddata
\end{deluxetable}

\begin{figure}
\begin{center}
\includegraphics[scale=0.6]{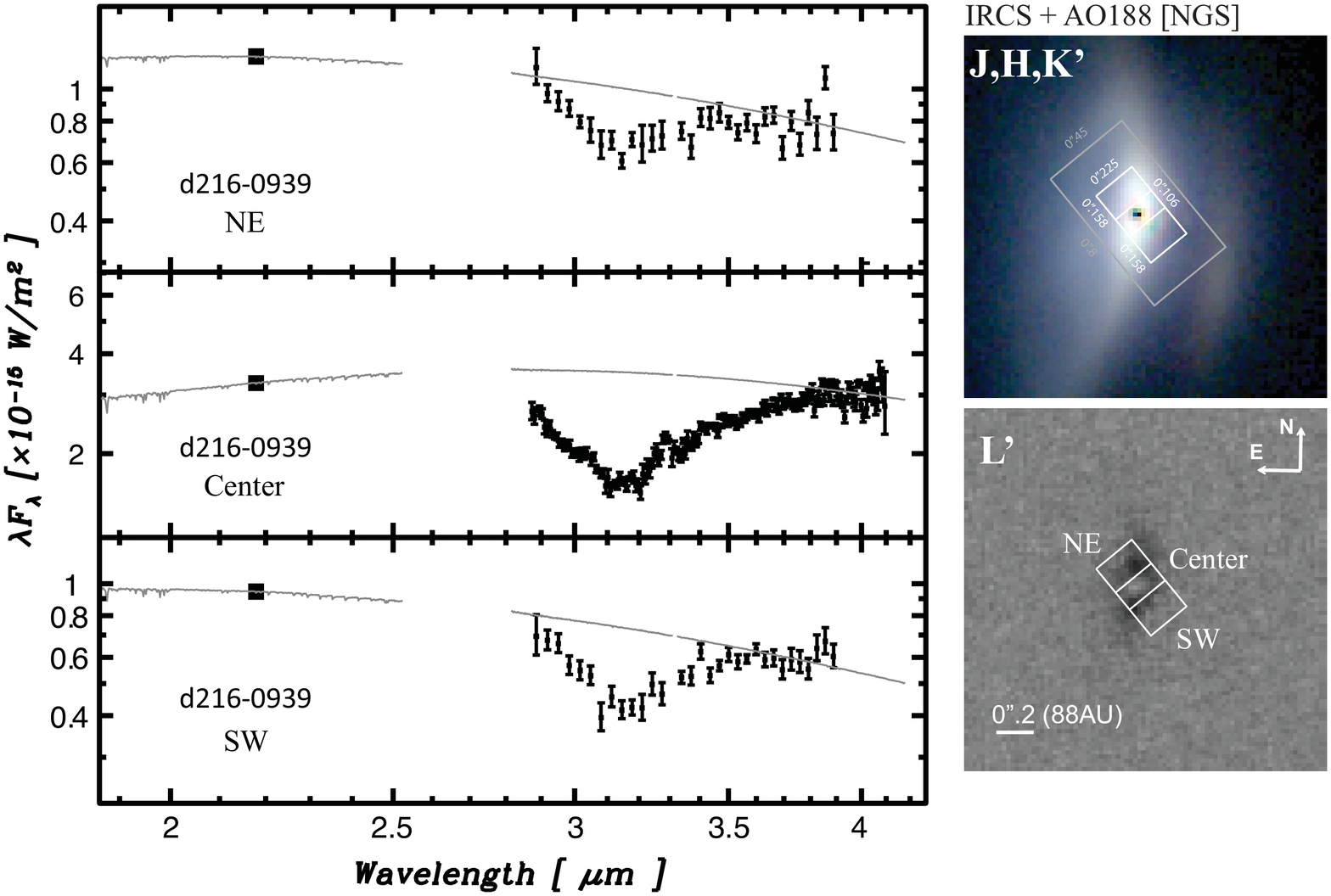}
\caption{\label{fig-resolved-bin}Spatially resolved $L$-band spectra of d216-0939. Filled square shows $K^{\prime}$ photometric point, and gray lines are the best-fit model spectra to derive the water ice band for each positions. Right panels show the slit location and the area for extraction of the spectra, which are superposed on $J$, $H$, $K^{\prime}$ color composite image (upper) and $L^{\prime}$ image (lower) from the subtracted images in Figure~\ref{fig-objimage}. Gray box shows the slit location used on the first epoch of 2006-01-15-UT.
}
\end{center}
\end{figure}

For the case of d216-0939, the detection of the apparently comparable absorption at 3.1\,$\micron$ band on different epoch (2006-01-15-UT) has been reported in Paper I, and this water ice band is found to be unchanged after a time interval of 3.63 years. Since d216-0939 is spatially resolved in the $L^{\prime}$ image, spectra are extracted for the three locations (northeast (NE), center, and southwest (SW) of d216-0939) shown in the left panel of Figure~\ref{fig-resolved-bin}. Using the PSF of the nearby star JW 670, contribution of the wing of the PSF at $L^{\prime}$ to NE and SW is estimated to be 12\% and 13\%, respectively. Thus, the spectra for NE and SW are affected by the contribution from the center location. The position angle of the slit was set to the direction from the object to the nearby star JW 671, and the slit is not aligned to the disk direction. Whereas the NE position covers the $L^{\prime}$ scattered light region effectively, the SW position is around the mid-plane area with some scattered light. The derived spectra for each position are shown in the left panel of Figure~\ref{fig-resolved-bin}. The photometric points at $K^{\prime}$ and $L^{\prime}$ are estimated from the flux density within the slit as measured from the AO images. Model spectra are determined using the $K^{\prime}$ photometric points and normalized spectra at $L$, and those are shown with gray lines in Figure~\ref{fig-resolved-bin}. Both the positions of NE and SW show bluer continuum, and the water ice absorption is definitely more shallow, and the optical depths, center wavelengths, and widths of these spatially resolved water ice features are listed in Table~\ref{tbl-resolve-tau}.

\begin{deluxetable}{l c c c}
	\tablecolumns{4}
	\tablewidth{0pc}
    \tablecaption{Spatially Resolved Water Ice Absorption in d216-0939\label{tbl-resolve-tau}}
	\tablehead{
      \colhead{Location}&\colhead{Peak}&\colhead{$\tau_{ice}$}&\colhead{$\Delta\nu$}\\        	
      \colhead{in d216-0939}&\colhead{($\micron$)}&\colhead{}&\colhead{(cm$^{-1}$)}
	}
      \startdata
     Northeast (NE)&3.13&0.45$\pm$0.02&302\\
      Center&3.10&0.78$\pm$0.03&479\\
     Southwest (SW)&3.08&0.57$\pm$0.04&369\\
      \enddata
\end{deluxetable}

\section{Discussion}

\subsection{Water Ice in Individual Disks}

While d053-717 and d141-1952 show no water ice feature at 3.1\,$\micron$ band, moderate water ice absorption is detected for d132-1832 and d216-0939. In the following, each 3\,$\micron$ spectrum is individually discussed. 

\subsubsection{d053-717}
Although the morphology of the silhouette disk for this target obtained by $HST$ indicates the nearly edge-on geometry \citep{smi05}, the visual extinction has been estimated to be very low \citep[$A_{V}$=0.43;][]{hil97} from the optical spectroscopy. However, scattered light from central source may dominate the flux in the optical, which lead to the underestimate of the $A_{V}$. The object is located at moderate distance from O and B stars, and the UV environment is not too harsh to significantly affect the water ice distribution in the disk. Thus, the fact that no water ice feature at 3.1\,$\micron$ is detected toward d053-717 is a direct evidence for this object to have no significant amount of foreground extinction. \citet{get05} reported huge hydrogen column density (log($N_{H}$)=22.7 cm$^{-2}$) right at the position of d053-717 using the $Chandra$ $X$-$ray$ $observatory$, and the X-ray source is likely embedded in the circumstellar disk. This column density is converted to the $A_{V}$ of 31.5~mag by using the relationship of $N_{H}$ = (1.59 $\times$ 10$^{21}$) $\times$ $A_{V}$ cm$^{-2}$ \citep{ima01}, which should produce significant optical depth of the water ice absorption at 3.1\,$\micron$. Therefore, it is concluded that the real source associated with the silhouette disk is located behind d053-717. In our AO images at $K^{\prime}$ and $L^{\prime}$, no other source than d053-717 is detected.

\subsubsection{d132-1832}
The water ice absorption exists at 3.1\,$\micron$ with an optical depth of 0.70 for this target. The equation between column density of water ice ($N(H_{2}O)$) and visual extinction ($A_{V}$) adopted in Paper I, $N(H_{2}O)$ = (1$\times$10$^{17}$) $\times$ ($A_{V}$$-$$A_{V;0}$) \citep{whi96,tei99} is used for estimate of $A_{V}$. First, the column density of the water ice toward d132-1832 is derived to be 1.58$\times$10$^{18}$ cm$^{-2}$ using $N(H_{2}O)=\tau_{H_{2}O}\centerdot\Delta\nu/(2.0\times10^{-16})$ \citep{dhe86}. The threshold visual extinction of $A_{V;0}$ ranges from 2~mag to 10~mag as an indicator of harshness of the UV field among various star forming clouds. As a result, $A_{V}$ for d132-1832 is estimated to be 17.6--25.6~mag. This $A_{V}$ is far greater than the average visual extinction (1--2~mag) of the foreground ``neutral lid'' in M42 \citep{ode92} and also significantly larger than the more local visual extinction ($\sim$10~mag) showed in Figure~\ref{fig-sample} \citep{sca11}. Therefore the detected water ice is concluded to be localized around the object d132-1832. Since the nearby star 2MASS J05351361-0518329 has no water ice absorption, there is no common water ice material in front of those two stars. Assuming that d132-1832 and 2MASS J05351361-0518329 are located within the same region of M42, then the detected water ice originates from the region within a projected distance of 5\farcs68 (2352 AU) around d132-1832. 

\begin{figure}
\begin{center}
\includegraphics[scale=0.6]{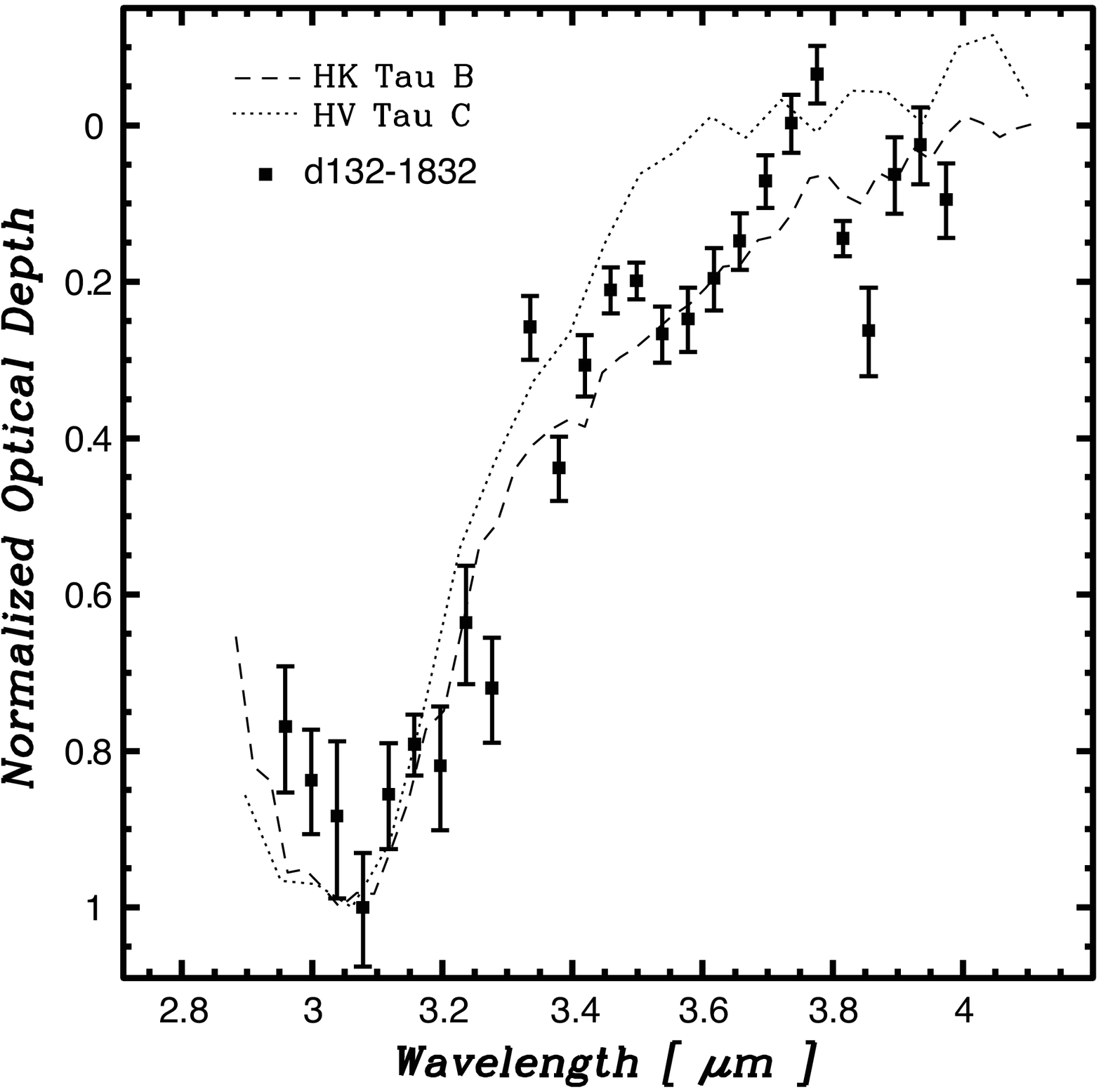}
\caption{\label{fig-d132-tau}Normalized optical depth of the water ice feature for d132-1832. In contrast to d216-0939, no 3.2\,$\micron$ enhancement of the optical depth is observed.
}
\end{center}
\end{figure}

Since the detailed analysis of the detected water ice profile for d132-1832 is somewhat difficult with the S/N obtained, the overall profile of the water ice feature extracted in Figure~\ref{fig-d132-tau} shows no broad absorption around 3.2\,$\micron$ and it is similar to the optical depth detected in the Taurus edge-on disks \citep{ter07}. This suggests that the detected water ice grains originate from regions further out in the circumstellar disk rather than in the inner disk where large crystallized ice grains can be produced by grain growth and thermal processing.

Interestingly, \citet{smi05} pointed out that d132-1832 has been variable in the images taken with H$\alpha$ filter of $HST$. The central star seems clearly visible right at the center of the silhouette disk on the first observing epoch of 1998-04-04-UT, and then it becomes invisible on the second epoch of 2004-01-18-UT. 
This variability suggests a change of the extinction in the disk between two epochs or an H$\alpha$ flare on the first epoch \citep{smi05}. The image on another epoch (2005-04-10-UT) provided as an online data by \citet{ric08} still shows a little emission from the central star in the F658N and strong emission in the F775W and F850LP filters. The images taken with the F775W and F850LP filters show evidence of scattered light. It is unknown whether the central star was optically visible or invisible on the epoch of our observation (2011-10-23-UT). At the present time we cannot rule out variability due to an H$\alpha$ flare and the amount of scattered light is not quantitatively understood. If the cause of the variability comes from a change in the extinction, then it would imply the water ice absorption would vary as well.


\subsubsection{d141-1952}
While the silhouette disk around d141-1952 is the smallest in size among our target samples, its disk mass derived from the submm measurement is the second most massive among the four targets (Table~\ref{tbl-sample}). Assuming the abundance ratio between the water ice and the total dust mass to be similar among  the protoplanetary disks, abundant water ice exists in the disk of d141-1925. The silhouette disk shows the lowest inclination angle of 55--60\arcdeg among the targets, and most likely our line of sight toward the central star does not pass through the circumstellar disk around the object.

\subsubsection{Spatially Resolved d216-0939}

As shown in Figure~\ref{fig-resolved-bin}, the d216-0939 spectra were spatially resolved with 0\farcs1 resolution using the AO188 system with NGS mode. The bluish continua seen in the spectra of NE and SW positions are consistent with the bluer scattered light in those regions. 
The contamination from the central star mentioned in Section 4.2 affects NE and SW by the optical depth of 0.37 and 0.51, respectively. Therefore, the significant fraction of the optical depth measured at the locations of NE and SW originates from the water ice absorption at the center position, and the water ice absorption at NE and SW is very small ($\tau_{ice}$=0.08 and 0.06, respectively). These shallower optical depths at the position of NE and SW indicate that the water ice materials attributed to the absorption detected in NE and SW are located at greater height from the mid-plane. 



\begin{figure}
\begin{center}
\includegraphics[scale=0.6]{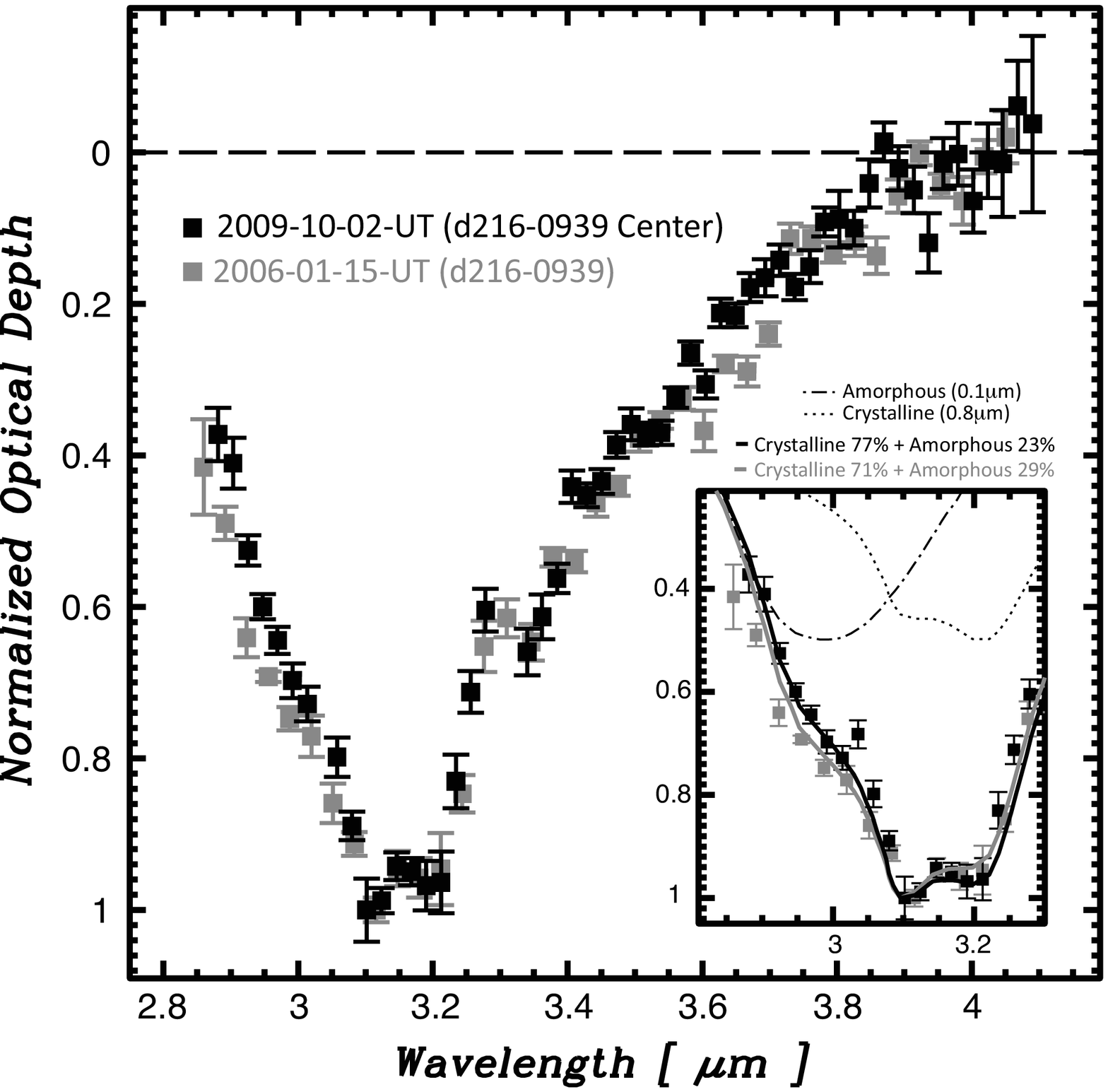}
\caption{\label{fig-d216-comp}Comparison of the normalized optical depth of the water ice absorption between large aperture (0\farcs45 $\times$ 0\farcs8) on 2006-01-15-UT and smaller region (0\farcs225 $\times$ 0\farcs11) on 2009-10-02-UT around the photo center of d216-0939.
}
\end{center}
\end{figure}

Focusing on the water ice feature detected in the center location of d216-0939, the profile of the water ice absorption is compared with the water ice profile reported in Paper I with a larger aperture (0\farcs45 $\times$ 0\farcs8) in Figure~\ref{fig-d216-comp}. In the figure, the profiles of the water ice absorption are found to be almost identical with a possible decrease of the optical depth around 2.9\,$\micron$ and 3.6\,$\micron$. The pure amorphous water ice has an absorption feature at 2.9\,$\micron$ , and therefore the decrease at 2.9\,$\micron$ can be explained by the lower fraction of the amorphous water ice. Although the difference of the profiles at 2.9\,$\micron$ is small, the water ice composition is quantitatively estimated in the following by assuming that the difference is real. We apply exactly the same method as used in Paper I, and the new result of the water ice optical depth in the region of 2.8--3.3\,$\micron$ is well matched to the composition of large-size crystallized water ice particles (77.2\%$\pm$4.0\%) and small-size amorphous water ice particles (22.8\%$\pm$4.0\%) with a minimum reduced $\chi^{2}$ and 1$\sigma$ uncertainty. Since the composition of the water ice detected on the first epoch is crystallized water ice particles of 71.4\%$\pm$2.7\% and amorphous water ice particles of 28.6\%$\pm$2.7\%, the results on those two epochs are overlapping within the 1$\sigma$ uncertainty. The optical depths of the crystallized water ice ($\tau_{crystal}$) are derived to be 0.55 and 0.69 on two observing epochs of 2006-01-15-UT and 2009-10-02-UT, respectively. Different apertures were used for spectroscopy on the two epochs, and therefore these derived optical depths cannot be directly compared. The large aperture of 0\farcs45 $\times$ 0\farcs8 used for extraction of the spectrum on the first epoch contains a large amount of scattered light from the central source. Since the large-size grains are thought to settle toward the mid-plane of the disk \citep[e.g.,][]{wil11}, the spectrum of the scattered light does not contain a significant amount of the large-size crystallized water ice. As a result, the scattered light component dilutes the optical depth of the crystallized water ice absorption. From the $L^{\prime}$ image in Figure~\ref{fig-objimage}, we estimate the fluxes of the scattered light component (with an aperture of 0\farcs45 $\times$ 0\farcs8) and the central region (with an aperture of 0\farcs225 $\times$ 0\farcs11) of d216-0939 to be 2.29 $\times$ 10$^{-16}$ and 8.22 $\times$ 10$^{-16}$ W m$^{-2}$ $\micron$$^{-1}$, respectively. 
If the scattered light is assumed to contain no large-size crystallized water ice feature, this dilution effect leads to the change of $\tau_{crystal}$ by 0.20. In reality, the scattered light includes at least partially the large-size crystallized water ice feature. Thus, the optical depth change due to the dilution effect is 0--0.20, and $\tau_{crystal}$ in the center region is estimated to be 0.55--0.75 on the first epoch. This optical depth range of the large-size crystallized water ice is consistent with $\tau_{crystal}$=0.69 detected on the second epoch.

\subsection{No Variation of Large-size Crystallized Water Ice in d216-0939 Disk}

The variability of the water ice features has been detected toward several young stellar objects \citep{lei01,bec04,bec07,ter07}. In particular, \citet{ter07} discovered the large variation ($\Delta\tau_{ice}$ $\sim$ 0.59) of the water ice in the edge-on disk around HV Tau C with a time interval of 2.32 years and derived the possible size of the water ice cloud to be 1.4~AU. No significant change of the large-size crystallized water ice feature in the time interval of 3.63 years indicates uniform distribution of the carrier of the water ice in the disk of d216-0939. Since the distribution of the crystallized water ice is not known well, we assume that it is located at the distance of 30~AU from the central star, which is suggested for the first time as the lower limit of the crystallized water ice distribution in YLW 16A \citep{sch10}. Based on this assumption, the possible water ice cloud  must be $>$ 3.48~AU in size.

\subsection{Critical Parameters for Water Ice Detection}
 
\begin{figure*}
\begin{center}
\includegraphics[scale=0.45]{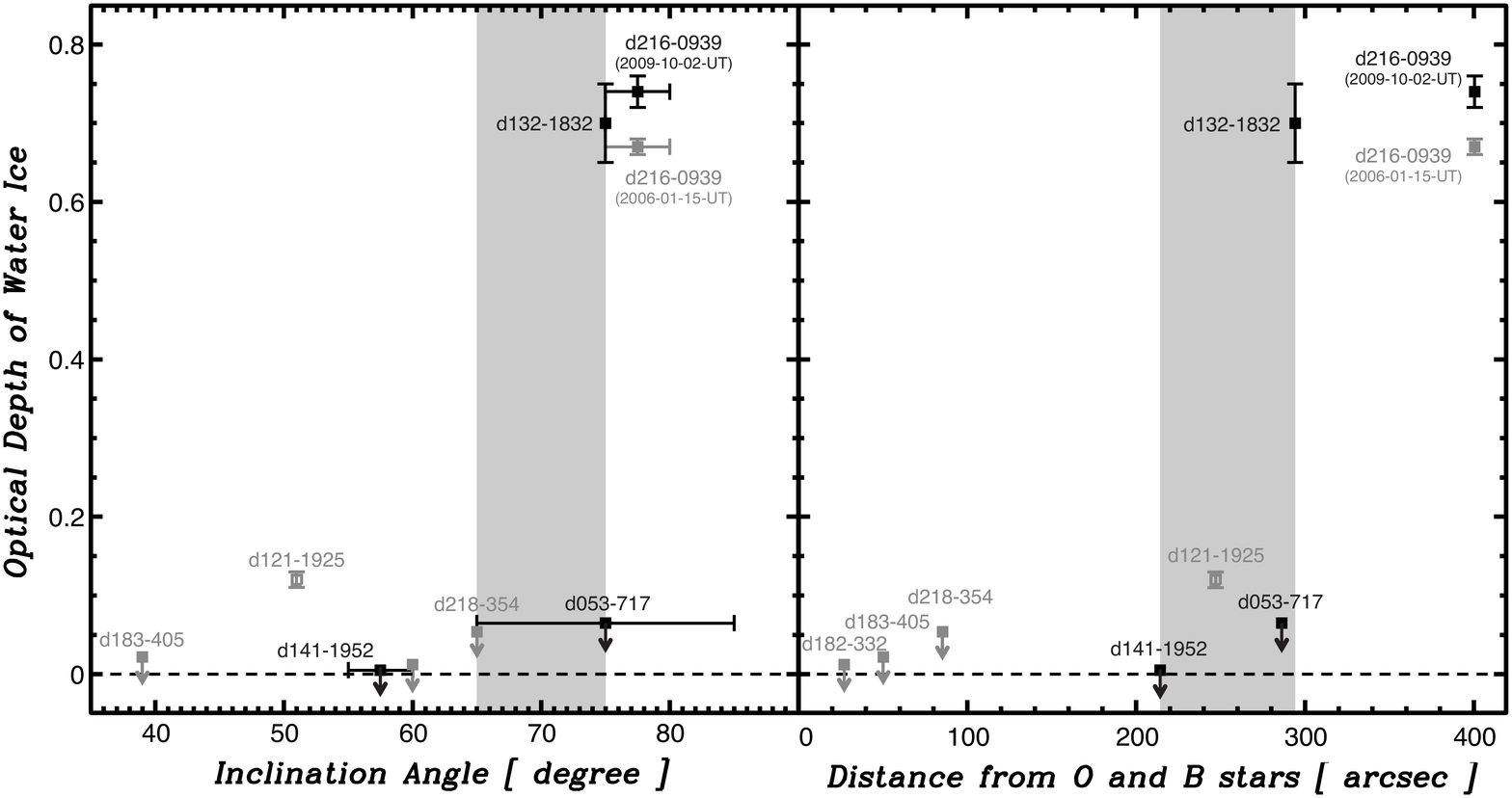}
\caption{\label{fig-para-ice}Relationship of the optical depth of the water ice absorption with the inclination angle of the disk (left panel) and the projected distance from O and B stars (right panel). Gray squares are from the result of Paper I, and black squares are reported here. The foreground water ice toward d121-1925 is presented with open gray square. A sudden change of the optical depth is seen around the shaded area in both panels.
}
\end{center}
\end{figure*}

Combined with the water ice features detected in Paper I, the relationship of the water ice optical depth with the inclination angle of the disks and the distance from the O and B stars is summarized in Figure~\ref{fig-para-ice}. It is noted that the water ice for d121-1925 is attributed to the foreground material. In addition, as pointed out in the above section, the lack of a water ice feature toward d053-717 is not well understood, but most likely d053-717 is not embedded in the silhouette disk. In this updated figure with the new samples, the rapid change of the optical depth can be found in the gray area for both the inclination angle of the disk and the distance from the O and B stars. Since UV radiation responsible for photodesorption of the water ice is gradually changed as a function of the distance, it is difficult to explain such a sudden change only by the difference of the distance for d141-1925 (214\farcs18) and d132-1832 (294\farcs19). On the other hand, as predicted in the model by \citep{pon05}, the different inclination angle can produce this kind of rapid change of the water ice absorption. Therefore, we conclude the primary parameter for explaining the detected water ice absorption to be the inclination angle of the disk, and its critical angle is confirmed to be 65--75$\arcdeg$. This suggests that the opening half-angles of the silhouette disks are around 70$\arcdeg$. This angle is almost the same angle derived for CRBR 2422-3423 in $\rho$ Ophiuchus by \citet{pon05}, and the intense UV radiation does not significantly affect the critical inclination angle of the disks.

\section{Summary}

We present the AO images and spectra of the four silhouette disks in the outskirt region of M42 and M43, and the following results are reported.

\begin{itemize}
\item[1.]{No water ice feature toward d053-717 and d141-1953 was detected, and moderate absorption of the water ice was found in the disks of d132-1832 and d216-0939.}
\item[2.]{The water ice profile of d132-1832 shows no signature of large-size crystallized water ice, and thus it may not have experienced grain growth in the disk.}
\item[3.]{Water ice absorption in the d216-0939 disk was spatially resolved. The center region shows the deepest absorption, and the large-size crystallized water ice component is identical to that detected on 2006-01-15-UT.}
\item[4.]{No time variation of the water ice in the d216-0939 disk was detected. Since the time interval is 3.63 years, this suggests a uniform distribution of the large-size crystallized water ice in the lateral direction of 3.48~AU.}
\item[5.]{Summarizing the optical depths of the water ice absorption in the silhouette disks, no significant water ice absorption is detected for the disks with an inclination angle of $<$ 65$\arcdeg$. Omitting the peculiar sample of d053-717 (see Section 5.1.1), there seems to be a critical inclination angle of the silhouette disks of 65--75$\arcdeg$ for water ice to be detected, which agrees to the model of \citet{pon05}. This implies that the opening half-angle of the silhouette disks is about 70$\arcdeg$.}
\end{itemize}

\acknowledgements
We thank all the Subaru telescope staff for maintaining the IRCS, AO36, and AO188 and supporting the observation. We also thank the referee for many helpful comments. We acknowledge the development efforts of the instrument to the development team members to realize these high sensitivity observations.

{}

\end{document}